\documentstyle[12pt]{article}    
\begin{document}
\begin{center}
{\bf Chronology Protection in Generalized G\"odel Spacetime }
{\vspace {3cm}}

             Wung-Hong Huang\\
             Department of Physics\\
              National Cheng Kung University\\
              Tainan, 70101, Taiwan\\

\end{center}
{\vspace {3cm}}

   The effective action of a free scalar field propagating in the generalized G\"odel spacetime is evaluated by the zeta-function regularization method.   From the result we show that the renormalized stress energy tensor may be divergent at the chronology horizon.   This gives a support to the chronology protection conjecture.

\vspace{5cm}
\begin{flushleft}

Classification Number: 04.62.+v; 04.20.Gz\\
E-mail: whhwung@mail.ncku.edu.tw\\
Physical Review D60 (1999) 067505

\end{flushleft}

{\newpage}

	Many people have considered the problem of time travel [1-13].  In recent, Hawking proposed the "chronology protection conjecture" [6] which states that the laws of physics will always prevent a spacetime to form the closed timelike curves (CTCs).   In his argument, the conjecture may be due to the divergence of the stress tensor near the chronology horizon where CTCs are beginning to form.   Several model spacetimes have been considered: the wormhole spacetime with a "time machine" [1-4], the Gott,s two-string spacetime[5], Grant space [7] , and the Misner spacetime [8,12-14].   In a number of papers it is concluded that the time machine is quantum unstable, as the stress tensor becomes divergent near the chronology horizon.   However, some authors [7-13]claim that the stress tensor is finite everywhere and it poses a problem for the chronology protection.

   For instance, despite of the divergent behavior shown in the first calculation by Hiscock and Konkowski [14], a specific-choice of the vacuum state in the Misner spacetime [11,13] could render the renormalized stress tensor vanish at the chronology horizon.   Boulware, Tanaka, and Hiscock [8] also find that for sufficiently massive fields in Gott space and Grant space, respectively, the renormalized stress tensor could remain regular at the chronology horizon.   In [13] it also argues that the renormalized stress tensor may be smoothed out by introducing the absorption material, so that the spacetime with a time machine may be stable against vacuum fluctuation.   The investigation of the Roman ring of traversable wormhole in [10] also find that the vacuum fluctuation can be made arbitrarily small.   However, in [12] it has argued that, even though a self-consistent vacuum state in the Misner spacetime could be constructed, it must ultimately be unstable against a slight perturbation.   Therefore, the chronology protection conjecture may be right.   Some people [6,10] have further argued that the solving the problem of chronology protection is impossible within the semiclassical theory of gravitation and it requires a fully developed theory of quantum gravity.   Anyway, the general proof of the chronology protection conjecture has not yet existed and it is useful to study the quantum stress tensor of a physical matter near a chronology horizon in various spacetimes with CTCs. 

   In this paper we will consider the generalized G\"odel spacetime [15]

	$$ds^2 = -dt^2 + dx^2 - 2 {\sqrt 2 }~ {\Omega\over \alpha}~ e^{\alpha x}~ dy dt + {1\over 2} (1 - 4{\Omega^2\over \alpha^2}) ~e^{2\alpha x}~ dy^2 + \alpha ^2dz^2 .   \eqno (1) $$
Note that we have introduced a constant metric $g_{zz} = a^2$  in the above equation as this could help us to find the renormalized stress tensor in the later calculation.   The case of $\alpha  = \sqrt 2~ \Omega$  is a rotating spacetime originally proposed by G\"odel [16] .   As analyzed in [15], there are no CTCs in the spacetimes with $\alpha^2 \geq\ 4\Omega^2$ .    But the spacetimes with $\alpha^2 < 4\Omega^2$  admit CTCs.   We will use the zeta-function regularization method [17] to evaluate the renormalized effective action of a free scalar field propagating in the generalized G\"odel spacetime.   The result shows that the action is divergent at chronology horizon , i.e. $\alpha^2 = 4\Omega^2$ .  From the result we can show that the stress energy may be divergent at the chronology horizon and this thus gives a new example to support the chronology protection conjecture. 

	In a system with a Hamiltonian  H  the renormalized effective action W evaluated by the zeta-function regularization method [17] can be expressed as 

$$		W = - {i\over 2}  [\zeta'(0) + \zeta(0) ln(\mu^2)],	 \eqno (2)$$
where 
$$		\zeta(\nu) = Tr(H)^{-\nu} . 	 			\eqno (3)$$
To evaluate the above zeta function exactly we shall first find the eigenvalue of the Hamiltonian H.   It is easy to see that the eigenvalue $\triangle$ of the Hamiltonian H for a massive (m) scalar field with curvature coupling ($\xi$) to the generalized G\"odel spacetime can be easily found through the same method [18,19] as that in the old G\"odel spacetime.    The method is that we first assume the following form of the eigenfunction 
$$ 		 \Psi(t,x,y,z) = h(x) exp(-i\omega t + ik_y y+ ik_z z) .  \eqno (4)$$
Then the function h(x) will satisfy the Whittaker's function and the condition to have an everywhere-bounded solution is that 

$$\triangle = a^{ -2}  k_z^2 + m^2 + \xi  R + ({\alpha\over 2})^2 + (4\Omega ^2 - \alpha^2)(n + {1\over 2} )^2 - [\omega - 2 \Omega \epsilon ( n + {1\over 2} )]^2,		 \eqno (5)$$
where the scalar curvature $R= 2(\alpha ^2 - \Omega^2)$ , $\epsilon = k_z /\mid k_z   \mid $  , and $n = 0,1,2,\cdot\cdot\cdot$. The zeta function can then be calculated by the formula [19]

$$ \zeta(\nu) = \int dx^4 \sqrt{-g} \int dk_z \int  d\omega \sum_{n=0}^\infty ~(\triangle)^{-\nu} .  \eqno (6)$$
Note that in [19] we have evaluated the finite-temperature effective potential for a scalar field in the old G\"odel spacetime, also with the help of zeta-function regularization method and thus the equation (6).  

 	Now, denoting $M^2 = m^2 + \xi R + (\alpha/2)^2 $ and rotating $\omega $ through $\pi/2$ in the complex plane to become  $i\omega_E $ [17] we have the relation  

$$  \zeta(\nu) = \int dx^4 \sqrt{-g} \int dk_z \int  d\omega_E [a^{-2}  k_z^2 + \omega_E^2 + M^2 + (4 \Omega^2 -\alpha^2)(n + {1\over 2})^2]^{-\nu } $$

$$= i a \pi (\nu-1)^{-1} \int dx^4 \sqrt{-g} \sum_{n=0}^\infty [M^2 + (4 \Omega^2 - \alpha^2) (n +{1\over 2})^2]^{1-\nu} .   \eqno (7)$$
Next, using the relation 

$$ \sum_{n=0}^\infty F[(n+1/2)^2] = \sum_{n=1}^\infty F[(n/2)^2]  -  \sum_{n=1}^\infty F[n^2] ,  \eqno (8)$$
in which F is an arbitrary function, we have

$$ \zeta(\nu) = i \pi a ~(\nu-1)^{-1}\int dx^4 \sqrt{-g}  \sum_{n=1}^\infty [M^2 + (4\Omega^2 - \alpha^2)(n /2 )^2]^{1-\nu}$$
$$ - i \pi a ~(\nu-1)^{-1}\int dx^4 \sqrt{-g}  \sum_{n=1}^\infty [M^2 + (4\Omega^2 - \alpha^2) n ^2]^{1-\nu}.     \eqno (9)$$
To proceed, let us quote a formula [20]

$$ \sum_{n=1}^ \infty (A n^2 + B)^{-s}  ={1\over 2} A^{-1/2} B^{1/2 -s} \pi \Gamma(s - 1/2) /\Gamma (s) + {1\over 2} (A + B)^{-s} - \int_0^1 (A x^2 + B)^{-s} dx +$$
$$ i \int_0^\infty [~[A(1+ix)^2 + B]^{-s} - [A(1-ix)^2 + B]^{-s}] (e^{2\pi x} - 1)^{-1} dx .  \eqno (10)$$
Then, after the calculations the zeta function becomes 

$$ \zeta(\nu) = i  \nu  \int dx^4 \sqrt{-g}  {7 \over 12}a \pi {1\over \sqrt{4 \Omega^2 - \alpha^2}}  [m^2 +\xi  R + (\alpha/2)^2 ]^{3/2}  + O(\nu^2) +$$
$$ \hspace{2cm} (finite ~ terms ~when ~ 4 \Omega ^2 ~approaches ~to~ \alpha^2) .    \eqno (11)$$
Using this result and from (2) we see that 

$$ 	 <T_{zz}> ={2 \over {\sqrt{-g}}}{\delta W\over \delta g_{zz}}   = {7 \over 12} \pi {1\over \sqrt{4 \Omega^2 - \alpha^2}}  [m^2 +\xi  R + (\alpha/2)^2 ]^{3/2} + $$
$$   \hspace{3cm} (finite ~ terms ~when ~ 4 \Omega ^2 ~approaches ~to~ \alpha^2)  ,     \eqno (12)$$
after letting $a = \mu = 1$.    We thus see that, at least, one component of the renormalized stress energy tensor is divergent at the chronology horizon and the investigation thus gives a new support to the chronology protection conjecture.   The evaluation of all components of the stress energy tensor is more difficult and left to be found.   

   Finally, let us mention that the spacetime considered in this paper dose not like those will form CTCs after the cosmological evolution [1-13].   Thus our result dose not directly relate to the chronology protection conjecture initially suggested therein.   However, as a geometry with the causality violation, such as G\"odel spacetime, can be taken as a part of a globally causal spacetime [21].   Thus our investigation dose suggest that the physical law of requiring the finite stress tensor will forbid a geometry with the causality violation to be enclosed in a region which maintains the required condition of global causality.   This means that the physical law will protect a would-be advanced civilization to create a noncausal geometry to violate the chronology protection conjecture.

\begin{enumerate}

\item  M. S. Morris, K. S. Thorne and U. turtsever, Phys. Rev. Lett. 61. 1446 (1988).
\item   V. P. Frolov and I. D. Novikov, Phys. Rev. D 42, 1057 (1990).
\item   V. P. Frolov, Phys. Rev. D 43, 3878 (1991).
\item   S. W. Kim and K. S. Thorne, Phys. Rev. D 43, 3929 (1991).
\item   J. R. Gott, Phys. Rev. Lett. 66. 1126 (1991).
\item   S. Hawking, Phys. Rev. D 46, 603 (1992).
\item   J. D. E. Grant, Phys. Rev. D 47, 2388 (1993).
\item   D. G. Boulware, Phys. Rev. D 46, 4421 (1992); T. Tanaka and W. A. Hiscock, Phys. Rev. D 52, 4503 (1995).
\item   G. Klinkhammer, Phys. Rev. D 46, 3388 (1992); S. V. Krasnikov, ibid. 54, 7322 (1996).
\item    M. Lyutikov, Phys. Rev. D 49, 4041 (1994); D. Laurence, ibid. 50, 4957 (1994); M. Visser, ibid. 55, 5212 (1997).
\item    S. V. Sushkov, Class. Quantum Grav. 12, 1685 (1995); ibid. 14 , 523 (1997).
\item    M. J. Cassidy, Class. Quantum Grav. 14 , 3031 (1997); ibid. 15 , 2257 (1998)
\item    L-X Li, Quantum Grav. 13 , 2563 (1996); L-X Li and J. R. Gott, Phys. Rev. Lett. 80. 2980 (1998).
\item    W. A. Hiscock and D. A. Konkowski, Phys. Rev. D 26, 1225 (1982).
\item    M. J. Reboucas and J. Tiomno, Phys. Rev. D 28, 1251 (1983) ; M. J. Reboucas and A. F. F. Teixeira, Phys. Rev. D 34, 2985 (1988).
\item    K. G\"odel, Rev. Mod. Phys. 21, 447 (1949).
\item    N. D. Birrell and P. C. W. Davies, Quantum Field in Curved Space (Cambridge University Press, Cambridge, England, 1982).
\item    B. Mashhoon, Phys. Rev. D 11, 2679 (1975); W. A. Hiscock, ibid. 26, 1497 (1982).
\item    W. H. Huang, Finit Temperature Cosmological Phase Transition in a Rotational Spacetimes,Class. Quantum Grav. 8, 1471 (1991);\\ 
W. H. Huang, One-loop Effective Action on Rotational Spacetimes: $\zeta$ - Function Regularization and Swinger Perturbative Expansion, Ann. Phys. 254, 69 (1997).

\item    J. Toms, Phys. Rev. D 21, 2805 (1980); L. H. Ford, ibid. 21, 933 (1980). 
\item    M. Novello and M. C. Motta da Silva, Phys. Rev. D 49, 825 (1994)

\end{enumerate}
\end{document}